\documentclass[superscriptaddress,twocolumn,bibnotes,amsmath,amssymb,aps,pra,floatfix]{revtex4-2}

\usepackage{mathtools,bm,graphicx,hyperref,microtype,braket,centernot}

\usepackage[toc,page,titletoc]{appendix}
\newcommand{\nocontentsline}[3]{}
\let\origcontentsline\addcontentsline
\newcommand\stoptoc{\let\addcontentsline\nocontentsline}
\newcommand\resumetoc{\let\addcontentsline\origcontentsline}

\usepackage{wasysym} 
\usepackage{siunitx} 
\usepackage{upgreek} 
\DeclareSIUnit{\gauss}{\ensuremath{\mathrm{G}}}
\DeclareSIUnit{\bohrradius}{\ensuremath{a_0}}

\input cyracc.def
\newfam\cyrfam
\font\tencyr=wncyr10
\font\sevencyr=wncyr7
\font\fivecyr=wncyr5
\def\cyr{\fam\cyrfam\tencyr\cyracc}
\textfont\cyrfam=\tencyr \scriptfont\cyrfam=\sevencyr
\scriptscriptfont\cyrfam=\fivecyr
\newcommand{\TD}{{\cyr D}} 

\renewcommand{\Re}{\operatorname{Re}}
\renewcommand{\Im}{\operatorname{Im}}
\newcommand{\upa}{\uparrow}
\newcommand{\dna}{\downarrow}
\newcommand{\br}{\bm{r}}
\newcommand{\bR}{\bm{R}}
\newcommand{\bp}{\bm{p}}
\newcommand{\bP}{\bm{P}}

\newcommand{\bq}{\bm{q}}

\newcommand{\bv}{\bm{v}}
\newcommand{\bG}{\bm{G}} 
\newcommand{\ttime}{\tau}

\begin{document}

\title{\texorpdfstring{Lattice Unitarity: Saturated Collisional Resistivity in Hubbard Metals}{}}

\author{Frank Corapi}
\thanks{These two authors contributed equally.}
\author{Robyn T. Learn}
\thanks{These two authors contributed equally.}
\author{Benjamin Driesen}
\affiliation{Department of Physics, University of Toronto, Toronto, Ontario, Canada M5S 1A7}
\author{Antoine Lefebvre} 
\author{Xavier Leyronas}
\affiliation{Laboratoire de Physique de l'Ecole Normale Sup\'erieure, ENS, Universit\'e PSL,CNRS, Sorbonne Universit\'e, Universit\'e Paris Cit\'e, F-75005 Paris, France}
\author{Fr\'ed\'eric Chevy}
\affiliation{Laboratoire de Physique de l'Ecole Normale Sup\'erieure, ENS, Universit\'e PSL,CNRS, Sorbonne Universit\'e, Universit\'e Paris Cit\'e, F-75005 Paris, France}
\affiliation{Institut Universitaire de France (IUF), 75005 Paris, France}
\author{Cora J. Fujiwara}
\affiliation{Department of Physics, University of Toronto, Toronto, Ontario, Canada M5S 1A7}
\affiliation{Department of Physics, Lehigh University, Bethlehem, Pennsylvania 18015, USA}
\author{Joseph H. Thywissen}
\affiliation{Department of Physics, University of Toronto, Toronto, Ontario, Canada M5S 1A7}

\date{\today}

\begin{abstract} 
We investigate the interaction-induced resistivity of ultracold fermions in a three-dimensional optical lattice. In situ observations of transport dynamics enable the determination of real and imaginary resistivity. In the strongly interacting metallic regime, we observe a striking saturation of the current-dissipation rate towards a value that is independent of the interaction strength. This phenomenon is quantitatively captured by a dissipation model that uses a renormalized two-body scattering matrix. 
We further measure the temperature dependence of resistivity in the strongly interacting limit and discuss the predicted asymptotic high-temperature behavior. Our results provide a clear microscopic understanding of bounded resistivity of low-density metals, thus providing a useful benchmark for studies of strongly correlated atomic and electronic systems. 
\end{abstract}
\maketitle 


{\em Introduction.} Control of two-body interactions has enabled ultracold gases to explore new physical phenomena, such as the BEC-to-BCS crossover, Efimov states, resonant $p$-wave interactions, and dipolar droplets. 
The $s$-wave scattering length between spin-up and spin-down fermions is conveniently tuned using magnetic Feshbach resonances \cite{Verhaar:1993,FeshbachReview}; at resonance and in free space, one finds a scale-invariant unitary regime \cite{ZwergerBook}. 
Feshbach tuning has also been applied to fermions in periodic potentials, typically in pursuit of equilibrium many-body phase transitions \cite{Bloch:2008gl,Gross:2017do,tarruell2019review} near half filling ($n \to 0.5$) and at low temperature ($T \lesssim t$, where $t$ is the tunneling energy, and $T$ is given in units of energy). 

Optical lattices present an opportunity to study the $T\gtrsim t$ regime of the Fermi-Hubbard model (HM) without the complications of phonons or structural changes that would be found in typical materials \cite{Perepelitsky:2016jg}. While engineered disorder has been used to study localization \cite{ColloquiumMBL:2019}, optical standing waves are easily disorder-free. 
Instead, atom-atom scattering plays the dominant role inhibiting mass transport \cite{Strohmaier:2007hw,Schneider:2012,Ronzheimer:2013,Anderson:2019}. 
A further simplicity of ultracold dilute samples is targeted loading of the lowest-energy band. This avoids dissipation via band relaxation 
and offers a testbed for studies of the the single-band HM \cite{HubbardReview}, in which the on-site $s$-wave interaction between a spin-up atom and a spin-down atom is defined as $U$. 
Prior studies of the two-body problem for atoms in an optical lattice have considered eigenstates of the $t\to 0$ limit \cite{Esslinger:2006,venu:2023,Busch:1998,Calarco:2005,Blume:2012,PengZhang:2020} and band effects on $U$ \cite{Buchler10,Carr:2012,Cui:2010}. Transport studies have typically focused on the strongly correlated regime near $n=0.5$ \cite{Bakr:2018,Zwierlein:2018,Xu:2019}. 

\begin{figure}[b!]
\includegraphics[width=\columnwidth]{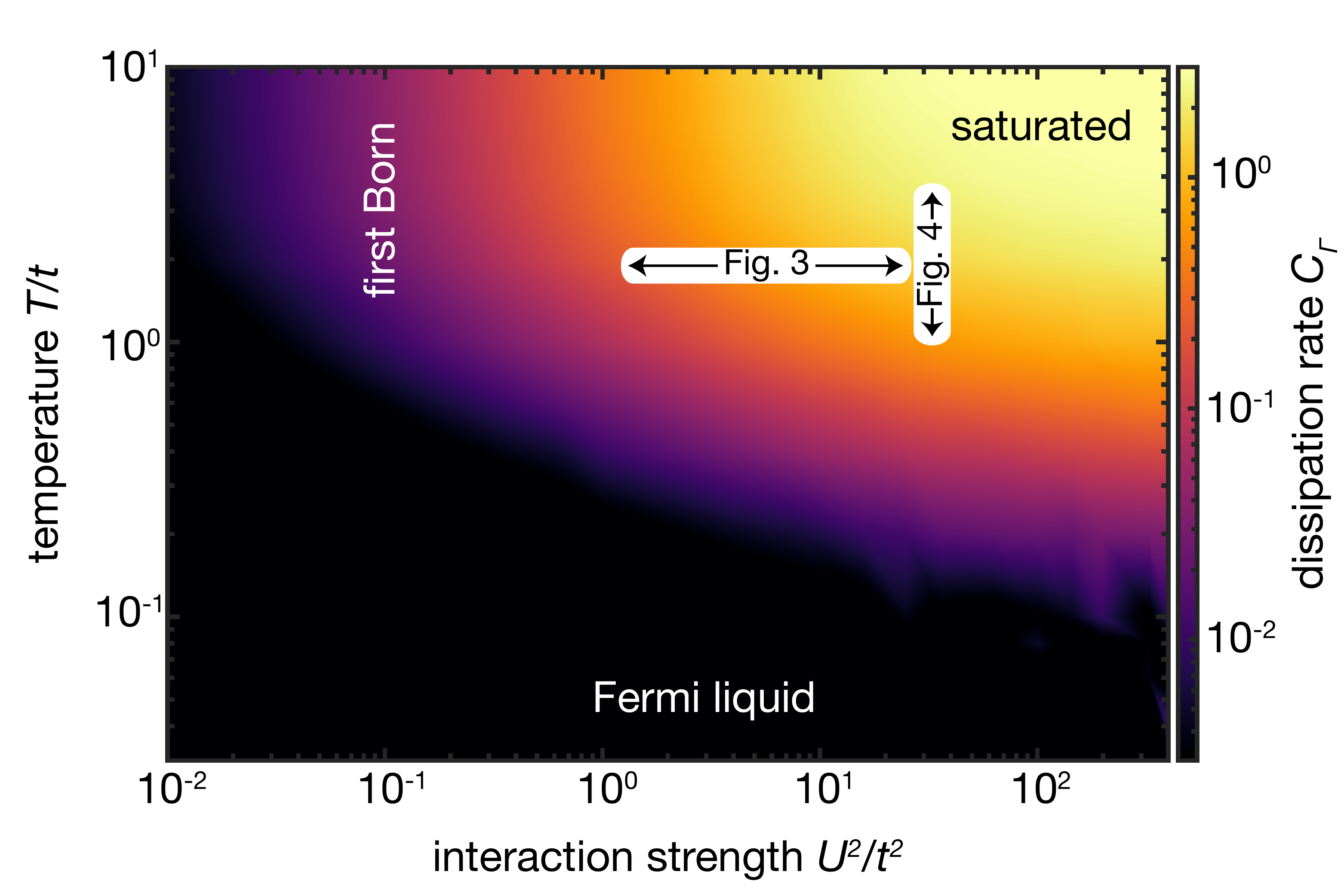}
\centering
\caption{\label{fig:regimes} {\bf Physical regime of measurements.} The current dissipation rate, normalized as $\hbar \Gamma/nt \equiv C_\Gamma$, is indicated by color throughout various regimes of temperature and repulsive on-site interaction strength $(U > 0)$, for average filling $n=0.035$. 
At low $T$, dissipation is described by the Fermi liquid picture. For $U^2 \lesssim t^2$, the first Born approximation of the scattering $\mathcal{T}$-matrix is valid. The measurements we report (indicated as ``Fig.\ 3'' and ``Fig.\ 4'') explore the regime beyond either approximation, in the crossover towards the saturated limit where $C_\Gamma$ is independent of $U$ and $T$.} 
\end{figure}

Here we study the resistivity of ultracold fermionic $^{40}$K in a three-dimensional cubic lattice in the $U \gtrsim t$, $T \gtrsim t$ regime. 
We work at low filling, $n \lesssim 0.1$, where the system is strongly interacting and yet weakly correlated, which enables a direct comparison to a non-perturbative solution of the two-body problem. 
We develop a dissipation model that connects the observed resistivity of a thermal ensemble to the microscopic collision rate. 
Figure~\ref{fig:regimes} shows the current dissipation rate $\Gamma$ anticipated for various $U$ and $T$, including a saturated regime that is the focus of our study. 
At the core of $\Gamma$ is the transition matrix $\mathcal{T}$ that gives the amplitude of Bloch-wave scattering. 
For $U^2 \ll t^2$, scattering is well described by the Born approximation at lowest order, i.e.\ $\mathcal{T} \approx U$. 
We find, both experimentally and theoretically, that in the $U^2 \gg t^2$ limit resistivity saturates towards a $U$-independent value. 
This observed saturation is distinguished from a saturation of static susceptibility through in situ calorimetry, and compares well to a Boltzmann calculation using a non-perturbative $\mathcal{T}$-matrix.

Just as in free space, where the unitarity of the scattering matrix prevents the cross-section from diverging when the scattering length becomes infinite, the scattering amplitude in a lattice remains bounded even in the limit of infinite $U$. 
Yet, contrary to the free-space case, the upper bound of the scattering cross-section cannot be reached in a thermal ensemble even when $U \to \infty$ due to the momentum dependence of the $\mathcal{T}$-matrix. We define this upper bound, where the $\mathcal{T}$-matrix is fully imaginary, as ``lattice unitarity''.
At our highest $U$, we infer that the dissipation rate reaches roughly one third of the lattice-unitary bound. 

\begin{figure}[tb!]
\centering
\includegraphics[width=\columnwidth]{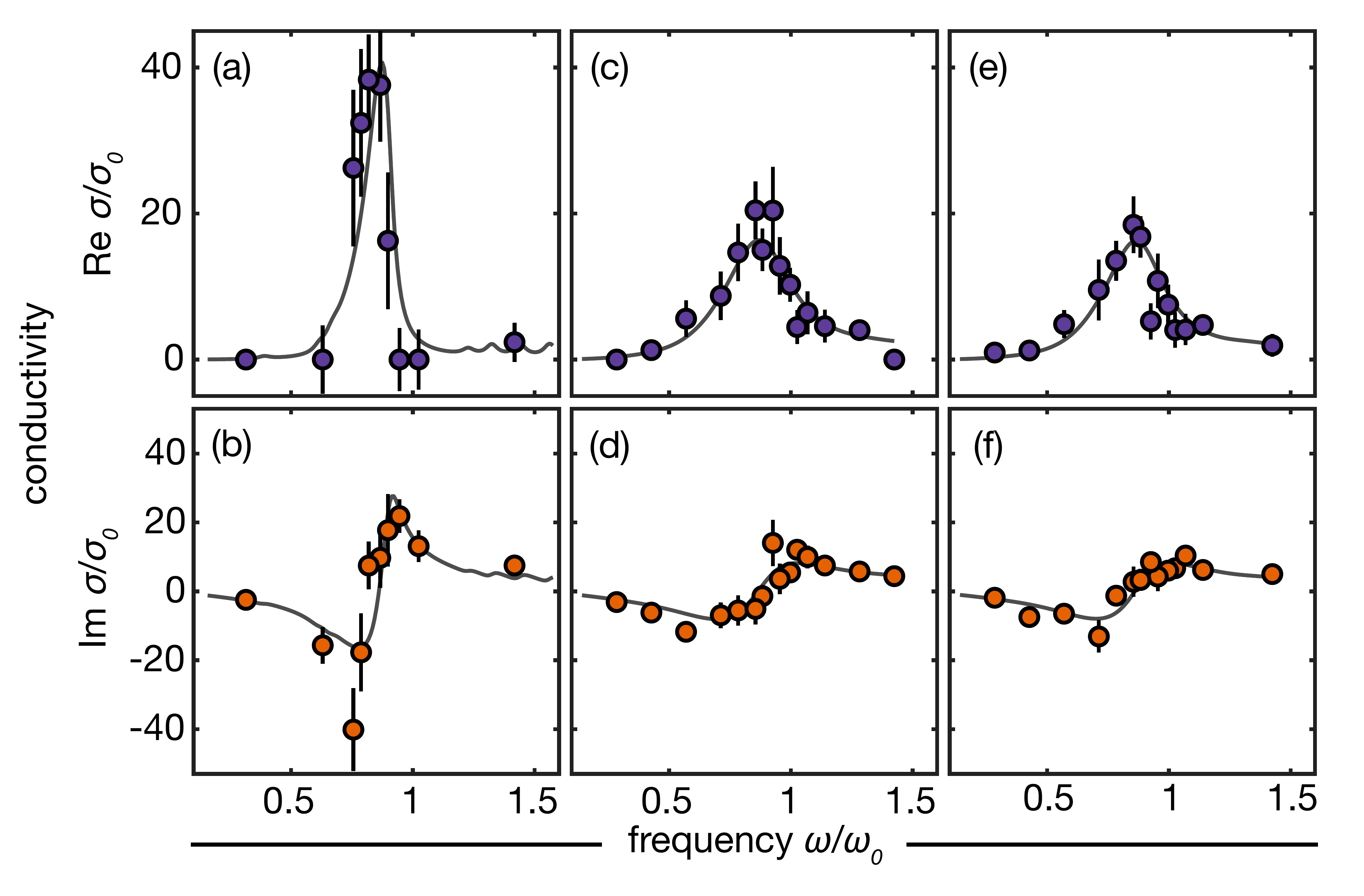}
\caption{\label{fig:saturation} {\bf Example of saturation.} 
The real and imaginary conductivity, scaled by $\sigma_0 = a_L^2 N/\hbar$, are shown versus drive frequency. Lines show a fit to a Kubo-type model, described in the text, which yield a best-fit width $\tau_Q^{-1}$. Spectra are shown at $U^2/t^2=1.15(11)$, $\tau_Q^{-1} = \qty{23(14)}{s^{-1}}$  (panels a,b), $U^2/t^2=16(2)$, $\tau_Q^{-1} = \qty{109(29)}{s^{-1}}$  (panels c,d), and $U^2/t^2=33(4)$, $\tau_Q^{-1} = \qty{110(19)}{s^{-1}}$ (panels e,f). Despite the strong increase in interactions, the broadening $\tau_Q^{-1}$ due to current dissipation saturates, with no distinguishable difference between c,d and e,f. Error bars indicate one-sigma uncertainty.} 
\end{figure}


{\em Measurement.} Our sample is a spin-balanced mixture of $N$ fermionic $^{40}$K atoms, trapped in a cubic lattice with period $a_L = \qty{0.53}{\um}$. 
Harmonic confinement, with isotropic oscillation frequency $\omega_0$ in the $xy$ plane, is created by two crossed optical dipole beams (XDT) and by the finite extent of the lattice beams. 
Measurements use a typical peak filling per spin state of $n_{\mathrm{pk}}=0.1$ and a lattice depth of $2.5(1)\,E_R$, where $E_R=h^2/(8 m a_L^2)$, for which the nearest-neighbor tunneling strength is $t/h = \qty{563(14)}{\hertz}$. Interactions are tuned by varying the $s$-wave scattering length $a_S$ with a magnetic Feshbach resonance near \qty{202}{\gauss}. Initial conditions are chosen by tuning the final trap-depth of evaporative cooling and by heating the sample (via short pulses of the lattice beams) prior to loading into the optical lattice. Both $T$ and $n_\mathrm{pk}$ are measured by comparing in situ filling to a thermal model \cite{SM}. 

We measure global conductivity $\sigma(\omega)$ through a technique proposed by Refs.~\cite{WZ14,WTZ15,TG11} and developed in Ref.~\cite{Anderson:2019}. 
Sinusoidal displacement of the XDT at frequency $\omega$ creates a uniform force amplitude $F_0 = m\omega_{\rm{XDT}}^2d_{\beta}$, proportional to the displacement $d_\beta$ along cartesian direction $\beta$, akin to an applied voltage in a charged system. Here $\omega_\mathrm{XDT}$ is the trap frequency due to the XDT beams alone along the driving direction. 
To allow a steady-state mass current to develop, $F_0$ is increased linearly over \qty{50}{\ms} and then held for an additional \qty{50}{\ms}, before studying two periods of the drive at  time intervals of $\pi/(4 \omega)$. We drive along the x-lattice such that $\beta=x$. 

The response of the cloud is measured via in situ fluorescence imaging of the central lattice plane \cite{SM}. The center of mass is fit to a sum of in-phase and out-of-phase sinusoids with respect to the drive: $R_x = S\sin(\omega \ttime)+C\cos(\omega \ttime)$, with free $\{S, C\}$. The real and imaginary components of $\sigma(\omega)$ are then $\Re[\sigma(\omega)] = N\omega C/F_0$ and $\Im[\sigma(\omega)] = N\omega S/F_0$. The complex resistivity $\rho(\omega)$ is simply $1/\sigma(\omega)$, yielding resistive $\Re[\rho(\omega)]$ and reactive $\Im[\rho(\omega)]$ components. 

Figure~\ref{fig:saturation} shows $\Re\sigma$ and $\Im\sigma$ across a range of $\omega$ for three different interaction strengths. 
Each spectrum shows a Drude-like response peaked near $\omega_0$ and a Kramers-Kronig dispersion in $\Im\sigma$ to accompany the resonance in $\Re\sigma$. 
At each frequency we choose $F_0$ both to remain in linear response \footnote{We find that a stringent test of linear response was to fit $\Re\sigma(\omega)$ and $\Im\sigma(\omega)$ independently, and then compare to the Kramers-Kronig relation.} and to control Joule heating \cite{JouleNote} such that the average temperature is relatively constant across the conductivity spectrum. We find that $|R_x| \lesssim \qty{1}{\um}$ is typically required to meet these constraints. 

The conductivity spectra are fit using a Kubo-type response function \cite{KuboSummary:2014,Anderson:2019} that uses the exact non-interacting basis states of the harmonically-confined lattice potential (see $\sigma_{xx}$ defined in the End Matter) and a relaxation-time approximation: that all eigenstates have a lifetime $\tau_\mathrm{Q}$. Examples are shown in Fig.~\ref{fig:saturation}. 
This common treatment~\cite{Michon:2023,AshcroftMermin,gotze:1972} captures the near-resonant broadening phenomenologically by treating $\tau_Q^{-1}$ as a fit parameter \footnote{In addition, we neglect any momentum- or frequency-dependence of $\tau_Q$, which leads to a violation of the DC susceptibility \cite{RTA2024Terada}. However, only near-resonant data has sufficient signal to contribute to $\Re \rho$.};  
a prediction for the current dissipation rate is the focus of the microscopic theory of transport described below. 

Conductivity $\sigma(\omega)$ is the Fourier transform of the current-current correlation function. The broadening observed between Fig.~\ref{fig:saturation}(a,b), with $U/t\approx 1$, to Fig.~\ref{fig:saturation}(c,d), with $U/t\approx 4$, arises from the reduced lifetime of currents due to scattering. However, one sees little change when further increasing to $U/t\approx 6$ (panels e,f). This is the phenomenon of saturation explored in our work. 
Perturbative scaling would predict a thirty-fold increase in current dissipation rate, proportional to $U^2$, while comparison of Figs.~\ref{fig:saturation}(a,b) and \ref{fig:saturation}(e,f) shows only a five-fold increase of $\tau_Q^{-1}$. 


{\em Dissipation model.} We interpret the observed phenomenon using a kinetic theory \cite{Abrikosov,Ziman:1956,Orso:2004,Schneider:2012,Kiely:2021}, which is expected to be valid for the weakly correlated metallic regime. 
The Boltzmann equation for the time evolution of the phase space distribution function $f(\bp,\br,\ttime)$ is
\begin{equation}
\partial_\ttime f+\bv_{\bp}\cdot\partial_{\br}f+{\bm F}\cdot \partial_{\bp}f=I_{\rm coll}[f]\label{eq:eqBoltzmantrap}
\end{equation}
where 
$\bv_{\bp}=\partial \varepsilon_{\bp}/\partial \bp$ is the group velocity, and the local force is 
${\bm F}=-m\,\omega_0^2\,\br+ F_0\,\cos(\omega \ttime)\,{\bm u}_x$. The collisional term, $I_{\rm coll}[f]$, represents the rate equation for the number of fermions in an arbitrary quasimomentum state, $\bp$, and is the sum of all scattering events into and out of that state. The scattering rate for a single event, $\Gamma_{12,34}$, at which two atoms in initial quasimomentum states $\{\bp_1,\bp_2\}$ scatter into $\{\bp_3,\bp_4\}$ can be calculated using the generalized Fermi golden rule and the exact two-body $\mathcal{T}$-matrix \cite{SM}.

For a single-band $d$-dimensional tight-binding model, the $\mathcal{T}$-matrix can be written as $\langle \bp_1,\bp_2|\hat{\mathcal{T}}|\bp_3,\bp_4\rangle =\mathcal{T}(\bP;E)/N_\mathrm{s}$, where $N_\mathrm{s}$ is the number of lattice sites, $\bP = \bp_1+\bp_2$ is the initial quasi-momentum, which is conserved up to a reciprocal lattice vector, and $E$ is the (conserved) total energy. 
Summing over all diagrams, one finds \cite{SM} $\mathcal{T}(\bP;E)^{-1}=U^{-1}-\TD(\bP,E+i 0^+)$, where $0^+$ is an infinitesimal positive quantity, and $\TD(\bP,z) = -i \int_0^\infty \!du \, e^{izu} \Pi_{\alpha=1}^{d} J_0[4 t \cos(a_L P_\alpha/2) u]$,  where $J_0(x)$ is a Bessel function of the first kind \footnote{This integral can be determined analytically in $d=2$; we use a numerical solution for $d=3$.}. For fixed momenta, an upper bound on the scattering amplitude is achieved in a lattice for $U^{-1}-\Re[\TD (\bP;E)]=0$, hence the lattice unitarity limit $\mathcal{T} \to -\Im[\TD]^{-1}$. 

In order to calculate $\sigma(\omega)$ and $\rho(\omega)$, we make use of the methods of moments \cite{SM}, which yields the exact equations
\begin{eqnarray}
\frac{d}{d\ttime}\langle r_{\alpha}\rangle&=&\langle  v_{\alpha}\rangle\label{eq:eqmotionx}\\
\frac{d}{d\ttime}\langle v_{\alpha}\rangle&=&\langle m^{-1}_{\alpha\beta}F_{\beta}\rangle+\frac{1}{N_{\sigma}}\int\!  v_{p_{\alpha}}I_\mathrm{coll}[f] \,. \label{eq:eqmotionv}
\end{eqnarray} 
To make further progress, we make the ansatz 
\begin{equation} \label{eq:fFDAnsatz}
f(\bp,\br,\ttime)=f^0[\bp-\bq(\tau),\br-\bR(\tau)]\,, \end{equation}
where $f^0=[e^{E(\bp,\br)/T-\mu/T}+1]^{-1}$ is the equilibrium Fermi-Dirac distribution, $\mu$ is the chemical potential, and  $E(\bp,\br)=\varepsilon_{\bp} + m \omega_0^2 r^2/2$ is the single-particle energy, including the trap potential. Here $\bq(\ttime)$ and $\bR(\ttime)$ are global shifts of momentum and position, respectively, of the equilibrium distribution. The ansatz is correct for the low-$\omega$ limit, where the response is purely in $\bR(\tau)$, and in the high-$\omega$ limit, where the response is purely in $\bq(\tau)$. Thanks to Kohn's theorem, the ansatz is also correct for arbitrary frequencies at low temperature and filling, since in this limit  we can approximate the single-particle Hamiltonian by that of a harmonic oscillator~\footnote{ In the absence of a lattice, the ansatz becomes asymptotically exact in the hydrodynamic limit, as well as for arbitrary collision rates in a harmonic trap. It is also commonly used in variational approaches, as in Ref.~\cite{Kiely:2021}.}. The ansatz does not support deformations of $f$ or fully capture edge-state behavior (see, for instance \cite{Pezze:2004}). We nonetheless adopt the ansatz as a minimal model, and find that it allows us to compute transport bounds. 

Assuming that the drive is small, we can expand the phase space density in $\bq(\ttime)$ and $\bR(\ttime)$: 
\begin{equation} f(\bp,\br,\ttime)=f^0(E)-\frac{\partial f^0}{\partial E}
\big[\bv\cdot\bq(\ttime)+m\,\omega_0^2\,\br\cdot\bR(\ttime)\big] + \ldots 
\label{eq:Ansatzline} \end{equation}
Evaluation of the terms in Eq.~\eqref{eq:eqmotionx} with this distribution function yields the following results \cite{SM}. First, 
$\langle r_{\alpha}\rangle=R_{\alpha}(\ttime)$ and 
$\langle v_{\alpha}\rangle=\langle m^{-1}_{\alpha\beta}\rangle_{\mathrm{eq}} \,q_{\beta}(\ttime)$, in which the expectation value of the effective-mass matrix is taken with the distribution at equilibrium $f^0$. For our isotropic lattice, $\langle  m^{-1}_{\alpha\beta}\rangle_{\mathrm{eq}} \equiv (m^*)^{-1} \delta_{\alpha,\beta}$, defining the effective mass $m^*$, which is a temperature- and density-dependent quantity, and whose low-energy limit tight-binding limit is $\hbar^2/2 t a_L^2$. The effective carrier number is $N m/m^*$. 

In the equation of motion for velocity, Eq.~\eqref{eq:eqmotionv}, the force term is $\langle m^{-1}_{\alpha\beta}F_{\beta}\rangle m^*=F_0 \delta_{\alpha,x}\cos(\omega \ttime)-m \omega_0^2\,R_{\alpha}(\ttime)$. The collisional term is 
$N_{\sigma}^{-1}\int\! v_{\alpha} I_\mathrm{coll}[f]=-B_{\alpha\beta}\,q_{\beta}$, where
\begin{eqnarray}
B_{\alpha\beta}&=&\frac{1}{N_{\sigma} T}
\iint\! d^3r \prod_{i=1}^{3}\frac{d^3p_i}{(2\pi\hbar)^3}\Gamma_{12,34}\,f_1^0\,f_2^0\,(1-f_3^0)\nonumber\\
&&
\times (1-f_4^0) v_{1,\alpha}(v_{1,\beta}+v_{2,\beta}-v_{3,\beta}-v_{4,\beta})\label{eq:Bint}
\end{eqnarray}
and $f_i^0 \equiv f^0(\bp_i,\br,\ttime)$. For an isotropic cubic lattice, $B_{\alpha\beta}=B\delta_{\alpha\beta}$. 
We define the quantity $\Gamma=B\,m^*$, which can be interpreted as the current dissipation rate. Since $\Gamma$ is linearly proportional to $n$ in the low-filling ($\mu \lesssim -4 t$) regime \cite{SM}, the dimensionless number $C_\Gamma = \hbar \Gamma/(n t)$ is a useful characterization of current dissipation. Figure~\ref{fig:regimes} shows $C_\Gamma$ calculated for a wide range of $T$ and $U$. 

The ansatz also leads to a set of differential equations for $R_{\alpha}(\ttime)$ and for $q_{\alpha}(\ttime)$. The equation for $R_{\alpha}(\ttime)$ is 
\begin{equation}
\ddot{R}_{\alpha}+\Gamma\,\dot{R}_{\alpha}+\frac{m}{m^*}\omega_0^2\,R_{\alpha}=\frac{F_0}{m^*}\delta_{\alpha,x}\cos(\omega \ttime)\,.
\label{eq:dampedosc}
\end{equation}
The stationary solutions of these damped-oscillator equations can be written $\Re(R_{\alpha,\omega}\,e^{-i\omega \ttime})$ and $\Re(q_{\alpha,\omega}\,e^{-i\omega \ttime})$. 
The complex amplitude of the particle current,  $J_{\alpha,\omega} =N(m^*)^{-1}\,q_{\alpha,\omega}$, gives the complex conductivity: 
$\sigma(\omega)=J_{\alpha,\omega}/F_0$ or $\rho(\omega)=\sigma(\omega)^{-1}$. 
One finds  
\begin{equation} \label{eq:resistivity} 
\rho(\omega) = \frac{m^*}{N}  \Gamma - i \frac{m^*}{N}  \omega + i \frac{m}{N} \omega_0^2\omega^{-1}
\end{equation} 
which is a Drude-like complex resistivity plus a capacitive impedance. The isolation of $\Gamma$ in the first term of Eq.~\eqref{eq:resistivity} underscores the logic of our measurement: $\Re \rho$ is a clean probe of collisional physics. 

\begin{figure*}[tb!]
\includegraphics[width=1.5\columnwidth]{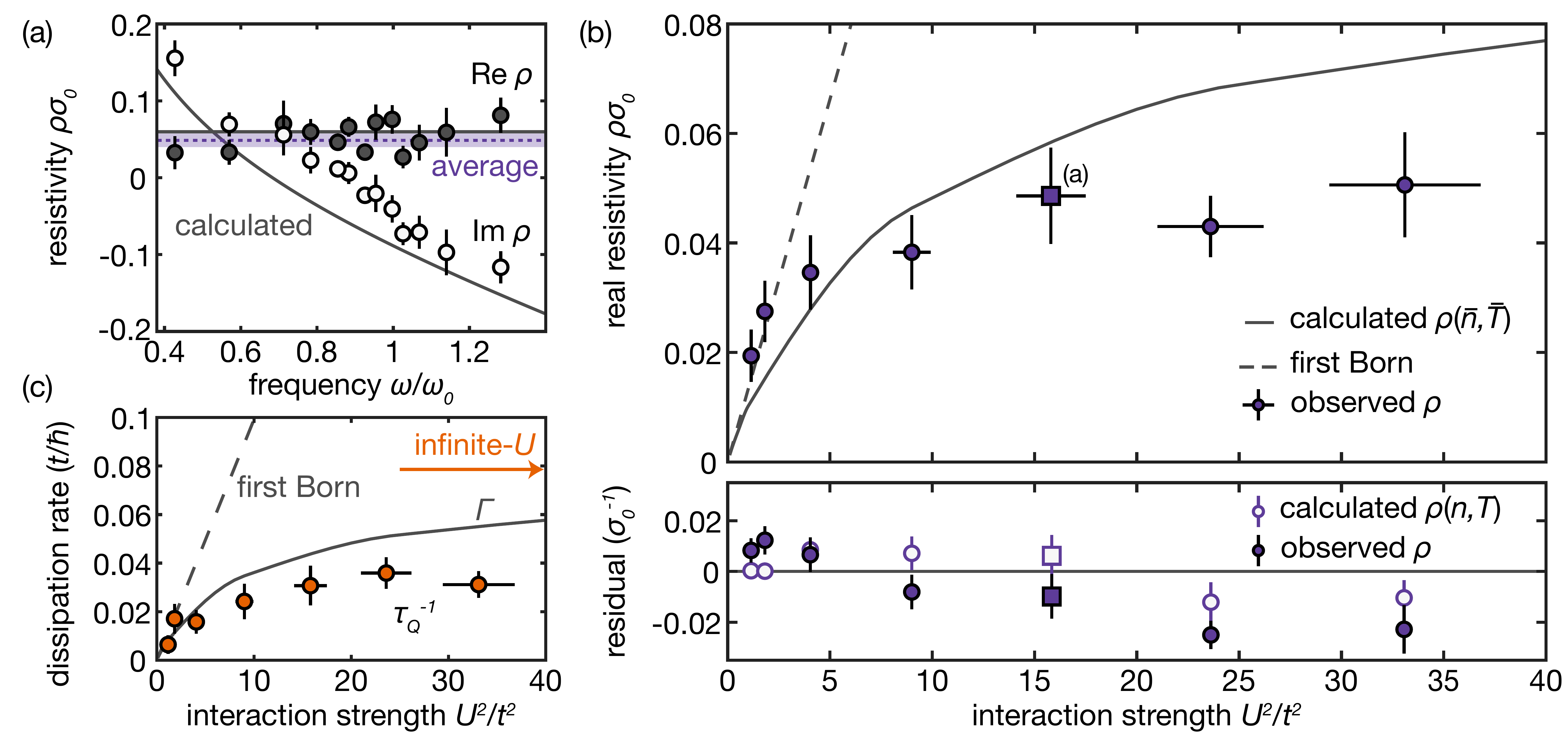}
\centering
\caption{\label{fig:Uscan} {\bf Resistivity saturation.} 
{\bf (a)}~The real and imaginary components of $\rho$ are shown versus drive frequency for $U^2/t^2=16(2)$, with one-sigma error bars shown at each $\omega$. While $\Im \rho$ decreases with $\omega$, $\Re \rho$ shows no clear trend. The weighted mean of $\Re\rho$ is indicated with a dotted horizontal line, and the 95\% confidence interval is shown by the shaded band. The solid lines show $\rho$ predicted by the dissipation model. 
{\bf (b)}~Frequency-averaged $\Re \rho$ data (points) are shown versus interaction strength $U^2/t^2$. 
The solid line shows a calculation of $\Gamma m^*/N$ using the full $\mathcal{T}$-matrix for the mean experimental conditions: $\bar T/t=2.19(18)$ and peak filling $\bar n_{\mathrm{pk}} = 0.09(1)$. The dashed line shows the first Born approximation, $\mathcal{T} = U$, for which $\rho$ would be proportional to $U^2$. Instead, $\Re\rho$ shows a clear saturation effect in $U$. The residual plot shows the difference between measured $\Re\rho$ (closed markers) and the predicted resistivity for $\bar T$ and $\bar n_{\mathrm{pk}}$. These residuals correlate with the predicted deviations of $\rho$ for the $T$ and $n_\mathrm{pk}$ of individual data sets (open makers). Error bars indicate twice the st.\ err.\ in $\rho$ from the ensemble of $T$ and $n_\mathrm{pk}$ across the $\sim 250$ measured clouds at each $U$. 
{\bf (c)}~Dissipation rate. $\hbar/(\tau_Q t)$ is determined through a fit to $\sigma(\omega)$ as in Fig.~\ref{fig:saturation}, and compared to the calculated $\hbar \Gamma/t$. The arrow shows the infinite-$U$ asymptote of the calculated $\hbar \Gamma/t$, and the dashed line is the first Born approximation.
}
\end{figure*}

Figure~\ref{fig:Uscan}(a) shows an example of the complex resistivity. 
In contrast with the resonance characteristic of $\Re\sigma$ (see Fig.~\ref{fig:saturation}), $\Re\rho$ does not have a peaked response. 
Both $\Re \rho$ and $\Im \rho$ can be compared to the dissipation model, whose predictions are shown as solid lines in Fig.~\ref{fig:Uscan}(a). 
The $\omega$-independent dissipative response agrees within error, and is discussed further below. The reactive response shows fair agreement with the prediction of Eq.~\eqref{eq:resistivity}, using the measured $T$ and $n_{\mathrm{pk}}$, and no free parameters. Refining our understanding of $\Im \rho$ at low frequency, such as measuring and calculating the deformations in $f(\bp,\br,\ttime)$ beyond Eq.~\eqref{eq:fFDAnsatz}, would be an interesting direction for future work. 

{\em Resistivity saturation.} 
Figure~\ref{fig:Uscan}(b) shows the frequency-averaged real resistivity \cite{SM}, measured at various $U/t$, from $U/t=1.07(5)$ to $U/t=5.75(33)$, at typical conditions $\bar{n}_{\mathrm{pk}} = 0.09(1)$ and $\bar{T} = 2.19(18)$. The solid line is the calculated real resistivity $\Gamma m^*/N$, at $\bar{T}$ and $\bar n_\mathrm{pk}$, using the full transition matrix without free fit parameters. We calculate $\Gamma$ and $m^*$ using a dispersion relation up to third order in tunneling \cite{SM}. For comparison, the dashed line shows the resistivity anticipated for the perturbative calculation $\mathcal{T} = U$, for which $\Re \rho$ scales as $n U^2/t^2$ at constant $T$ \cite{Anderson:2019}, with $n$ the volume-averaged filling~\cite{SM}.
In the strongly interacting regime, the measured resistivity shows a strong deviation from this first-Born approximation --- by as much as a factor of seven --- and is instead well described using the full $\mathcal{T}$-matrix. 
Note that the appearance of earlier saturation of the measured $\rho$ is primarily due to deviations in temperature and filling from the mean experimental conditions. This is apparent by comparing the data to calculations that use the $n_\mathrm{pk}$ and $T$ of each data set, instead of $\bar n_\mathrm{pk}$ and $\bar T$ [see residual plot in Fig.~\ref{fig:Uscan}(b)]. 
Comparison of the calculated $\Gamma$ to best-fit $\tau_Q^{-1}$ [Fig.~\ref{fig:Uscan}(c)] shows the same trend, and emphasizes that collisional resistivity saturation is a dynamical phenomenon, i.e., due to $\Gamma$ and not $m^*$. 
The observation and explanation of resistivity saturation is the primary result of our work. 

At infinite $U/t$ and low $n$, the dissipation model predicts the saturation of the current dissipation rate to a $U$-independent value $\Gamma = nt C_\Gamma(U\to\infty,T/t)/\hbar$. $C_\Gamma$, shown in Fig.~\ref{fig:regimes} for finite $U$ and $T$, gives the efficiency of current dissipation per scattering event, while $nt$ gives the inter-site collision rate. For $T=2.19t$, the dissipation model predicts $C_\Gamma(\infty,T/t)=2.527(5)$ (arrow in Fig.~\ref{fig:Uscan}(c)); at finite $U$ we measure up to $C_\Gamma =1.3(5)$, roughly half of the saturated value.   
The saturation of scattering can be understood with the approximation $\mathcal{T}(\bP,E) \approx \mathcal{T}(\bm{0},0)$, for which $\mathcal{T}^{-1} \approx U^{-1} + \xi_\infty/t$, with $\xi_\infty \approx i 0.22$. One then expects $|\mathcal{T}|^2$ to be half of its infinite-$U$ value at $U^2/t^2 \approx 20$, qualitatively similar to what is observed. 

We note that resistivity saturation at large $U/t$ is distinct from the saturation of the effective $U$ as a function of $a_S$ discussed in \cite{Buchler10,Cui:2010,Carr:2012}. Our data investigates only the regime $a_S \leq 0.14 a_L$, where we expect a reduction in $U$ from the linear-in-$a_S$ calculation to be $\lesssim 20 \%$.
Instead, the saturation phenomenon observed here is a dynamical effect: the saturation of the scattering amplitude. 

Contrary to the free-space case where unitarity is achieved at infinite $a_S$ for all momenta, the unitary condition becomes momentum-dependent in a lattice. Therefore, the damping rate does not reach the unitary limit, even when $U\rightarrow\infty$, due to the averaging over the particle momentum distribution. For $T/t=2.19$, replacing $\mathcal{T}$ by $|\Im \TD|^{-1}$ predicts a unitary $C_\Gamma = 4.16(3)$, which is about three times the largest measured $C_\Gamma$ at finite $U$, discussed above.

\begin{figure}[tb!] 
\includegraphics[width=\columnwidth]{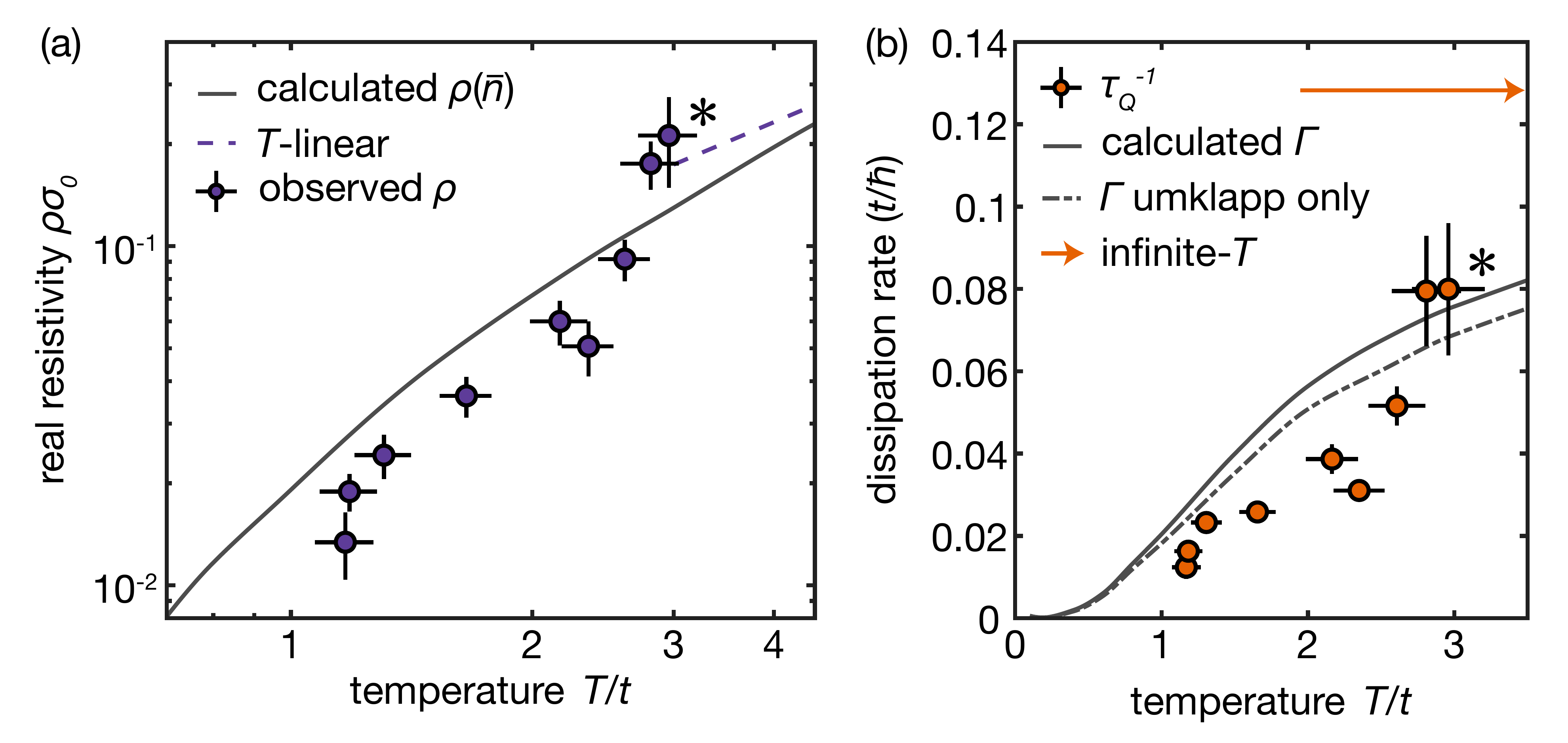}
\centering 
\caption{\label{fig:Tscan}{\bf Temperature dependence of interaction-saturated resistivity.} 
{\bf (a)}~Measured (points) and calculated (solid line) real resistivity are both determined as in Fig.~\ref{fig:Uscan}(b), here for $\bar n_{\mathrm{pk}} = 0.092(13)$, fixed $U/t=5.75(33)$, and variable $T$. 
The range of $T/t$ is expanded for comparison to the $T$-linear trend (dashed line) at higher temperature. 
{\bf (b)}~ Dissipation rate is determined as in Fig.~\ref{fig:Uscan}(c). $\hbar/(\tau_Q t)$ is compared to calculations of $\hbar \Gamma/t$ either with all events (solid line) or with only umklapp events (dash-dotted line). The arrow shows the infinite-$T$ asymptote of the calculated $\hbar \Gamma/t$. 
In both plots, the ``$\ast$'' flags a systematic issue with the two highest-$T$ points (see text and End Matter).
}
\end{figure}

The effect of temperature in the $U^2 \gg t^2$ regime is measured by preparing samples at various $T$, fixed $U$, and typical $\bar n_\mathrm{pk} = 0.092(13)$. $\Re\rho$ and $\tau_Q^{-1}$ are determined as above, using a complete spectral response at each $T$. 
Figure~\ref{fig:Tscan} shows these data and the results of dissipation-model calculations using the full $\mathcal{T}$-matrix. 
As anticipated by the model, the overall trend is a monotonic increase in $\Re \rho$, which can be attributed to an increase in both $\Gamma$ and $m^*$ at higher $T$. The impact of variable $T$ is strong: $\Re \rho$ increases by a factor of ten across the range accessible to our experiment. 
The two highest-$T$ data points shown are near our signal-to-noise limit, as we cannot measure resistivity reliably beyond $\rho\sigma_0\sim 0.1$ (see End Matter).
Overall, the satisfying agreement between measurements and the dissipation model, using no free parameters, demonstrates that a kinetic model with the full $\mathcal{T}$-matrix captures the temperature dependence of saturated collisional resistivity.

Figure~\ref{fig:Tscan}(b) isolates the dissipation rate, and compares calculated $\Gamma$ to the best-fit $\tau_Q^{-1}$. 
An alternate calculation uses only the umklapp events: scattering events in the collisional integral with a non-zero $\bm{G}$, i.e., where momentum is transferred from the lattice to an interacting pair. This is shown as a dotted-dashed line in Fig.~\ref{fig:Tscan}(b), from which we see that resistivity in the regime explored here is dominated by umklapp events. 

Let us consider next the $T \gg t$ limit, beyond the range of our data, while still restricting occupation to the lowest band. 
The band becomes uniformly filled, causing the current dissipation rate to saturate [see Fig.~\ref{fig:regimes} and Fig.~\ref{fig:Tscan}(b)]. The same is not true for $m^*$, which varies linearly with $T$ at high temperatures~\cite{Anderson:2019}. Therefore, as suggested by the dissipation model calculations shown in Fig.~\ref{fig:Tscan}(a), we expect $\Re \rho$ to increase linearly with temperature, with no signs of saturation, even beyond the uniform filling of the first band \cite{Perepelitsky:2016jg}. As observed by Brown {\it et al.}~\cite{Bakr:2018}, this is a purely thermodynamic effect. 

In the high-$T$ regime, the total collision rate $\gamma_\mathrm{coll}$ approaches $\Gamma$ \cite{SM}, so we can consider the interpretation $\gamma_\mathrm{coll} \approx n a_L^{-3} \sigma_\mathrm{coll} \bar{v}$, with cross section $\sigma_\mathrm{coll}$ and relative velocity $\bar{v}$. The latter approaches $t a_L/\hbar$ in the $T\gg t$ limit. This gives two physical insights into the regime we explore: 
first, that the mean free path is $\sim C_\Gamma^{-1} a_L/n$, which is $\gg a_L$ for the fillings we explore. (In other words, the $T$-linear scaling of $\rho$ expected here is not a ``bad metal''.) 
Second, the effective cross section is $\sim C_\Gamma a_L^2$. This is a striking quantum effect, since two neutral atoms scatter with a cross section that is $10^5$ times higher than their charge radii would suggest. 


{\em Conclusion.} 
In sum, we have observed the saturation of current dissipation in a strongly interacting system. 
The saturation phenomenon reflects a qualitative change in the nature of the scattering rate --- a crossover from interaction-limited to tunneling-limited dissipation. 
In this way, the observed phenomenon is conceptually similar to saturation of diffusivity \cite{enss2019universal} and shear viscosity \cite{Schafer:2009vf}, for example.
We have discussed the relationship between the saturation in damping rate and the unitary bound expected for collisions of Bloch waves. Whereas in free space the unitary limit is found for large scattering lengths, in a lattice the complexity of the dispersion relation introduces a dependence on the center-of-mass energy and momentum.
Our work provides a rare example of a resistivity measurement that is quantitatively compared to a first-principles calculation. Possible extensions could 
test the two-body model in the strongly correlated regime at higher densities, apply these methods to lower dimensions \cite{Schneider:2012,Ronzheimer:2013}, explore the emergence of hydrodynamics in Hubbard systems \cite{Kivelson:2011,Bakr:2018,Vucicevic:2023}, or probe the resistivity of a dilute low-temperature Fermi liquid, where umklapp scattering events are forbidden \cite{Rosch:2006gz,Maslov:2012,Kiely:2021}. 

\stoptoc

\begin{acknowledgments}
The authors would like to thank Rhys Anderson, Vijin Venu, and Peihang Xu for early work on this project, and Peng Zhang for sharing calculations on the two-body problem in a lattice. 
The authors acknowledge support from NSERC, AFOSR FA9550-24-1-0331, Institut Universitaire de France, PEPR Project Dyn-1D (ANR-23-PETQ-001), ANR Collaborative project ANR-25-CE47-6400 LowDCertif, and ANR International Project ANR-24-CE97-0007 QSOFT. 
The data that support the findings of this article are openly available \cite{DataBorealis}.  
\end{acknowledgments}

\bibliography{conductivity}

\begin{thebibliography}{51}
\bibitem [{Jou()}]{JouleNote}%
  \BibitemOpen
  \href@noop {} {}\bibinfo {note} {As suggested by \cite{TG11}, Joule heating is an alternate observable for determination of real conductivity (only), but not used in this work.}\BibitemShut {Stop}%
\end{thebibliography}


\onecolumngrid 
\begin{center}
\vspace{\columnsep}
\textbf{End Matter}
\vspace{0.25\columnsep}
\end{center}
\twocolumngrid

{\em Kubo fit model.} The lines in Fig.~\ref{fig:saturation}, and measured $\tau_Q$ in Figs.~\ref{fig:Uscan}(c) and \ref{fig:Tscan}(b), are based on linear-response theory for non-interacting atoms, with a phenomenological broadening $\tau_Q^{-1}$: 
\begin{eqnarray}\label{eq:Kubo}
\sigma_{xx}(\omega;\tau_Q)&&=\frac{N i \omega}{\hbar}\sum_{p\neq p'}\frac{(f_{p'}-f_p)|\bra{p'}\hat{R}_x\ket{p}|^2}{\omega-\omega_{pp'}+i\tau_Q^{-1}/2} \nonumber\\
&&
= \rho_{xx}^{-1}(\omega;\tau_Q),\label{eq:TDPT}
\end{eqnarray} 
where $\ket{p}$ are eigenstates of a lattice with overall harmonic confinement, $f_p$ are the occupation numbers, and $\hbar \omega_{pp'} = E_p - E_{p'}$ are the energetic splittings between states $\ket{p}$ and $\ket{p'}$. The inclusion of a state-independent $\tau_Q$ is a relaxation-time approximation. The subscript $xx$ denotes a response along the $x$-direction for a drive along the same direction. At the lattice depth used in this work, higher-order tunneling provides a measurable correction to the spectra. Energies and dipole-matrix elements are calculated by numerical diagonalization up to 15th order in tunneling. Since we work at low chemical potential, 
occupation numbers $f_p(T)$ are well approximated by Maxwell-Boltzmann statistics; corrections to $\sigma_{xx}$ due to Fermi-Dirac statistics ($\lesssim 4\%$) are equal to or below our typical measurement uncertainty. The first excited-band eigenstate is $\sim 14 t$ above the bottom of the ground band, so we restrict our treatment to the ground-band basis set. For $T \lesssim 3t$, the equilibrium population of the first excited band is $\lesssim 1\%$. 

While the harmonic confinement is ideally fixed, we find that pointing drift in the trapping beams induces small shifts in harmonic confinement at the $\pm1$\,Hz level. Therefore $\omega_0$ is also a free fit parameter for $\sigma$ spectra, (except for the two highest-$T$ data points, see below), and the eigenstates are recalculated during the fit runtime. All error bars derived from this linear-response model are obtained via bootstrapping with replacement. 

The relaxation-time approximation used here is that $\tau_Q$ is assumed to be independent of $p$ or $\omega$. However, $\Re \rho$ is not entirely frequency-independent for a finite system: $\Re\rho$ in Eq.~\eqref{eq:TDPT} shows a peak around $3\omega_0$ due to eigenstate anharmonicity and a plateau for $\omega>3\omega_0$ distinct from the low-frequency regime. One finds the ratio of $\Re\rho$ in the low-frequency regime ($\omega < 1.5 \omega_0$) to that in the high-frequency regime ($\omega > 10 \omega_0$) is $\sim 1.3$ (near saturation), where the value of $\Re\rho$ at high frequency approaches the ratio of $\tau_Q^{-1}$ to the spectral weight, $\frac{2}{\pi}\int_0^\infty\Re\sigma(\omega)d\omega$. 

{\em Low-signal analysis variant.} The conductivity spectra for the two highest-$T$ data points in Fig.~\ref{fig:Tscan} are broad and have a low signal-to-noise ratio. This has a twofold effect: we could not verify the Kramers-Kronig relation, and we needed to modify our fitting procedure to find $\tau_Q^{-1}$. For these spectra we could not reliably fit $\sigma_{xx}$ with a free $\omega_0$ or the measured $T$. We instead fixed $\omega_0$ to the calibrated value and allow $T$ to be a free fit parameter. We find that the fit temperatures, $3.6(6)t$ and $4.6(9)t$, overestimate the measured temperatures by $30\%$ and $50\%$, respectively. This suggests that these spectra were driven beyond linear response in an attempt for improved signal, in which case both $\tau_Q^{-1}$ and $\rho$ would be overestimated. These data are flagged with an asterisk in Fig.~\ref{fig:Tscan}. 

{\em Dissipation model conductivity.} From Eq.~\eqref{eq:dampedosc} we find a conductivity 
\begin{equation} \label{eq:conductivity} 
\sigma(\omega) = \frac{Ni\omega}{m^*}\frac{1}{\omega^2 - (m/m^*)\omega_0^2 + i\omega\Gamma}.
\end{equation} 
To compare this to the Kubo model, we use partial fraction decomposition to find a similar form:
\begin{eqnarray}
    \sigma(\omega) &=& \frac{Ni\omega}{2m^*\sqrt{m/m^*}\omega_0}
    \Bigg(\frac{1}{\omega - \sqrt{m/m^*}\omega_0 + i\Gamma/2}\nonumber\\
    &&
    - \frac{1}{\omega + \sqrt{m/m^*}\omega_0 + i\Gamma/2}\Bigg).
\end{eqnarray}
This form is equal to Eq.~\eqref{eq:conductivity} when the current damping rate is sufficiently low such that we can neglect $\Gamma^2$ terms. We can further simplify the summation by explicitly writing $m^*$ in the form calculated for Eq.~\eqref{eq:Kubo} through the sum rule, finding
\begin{equation}
\frac{1}{m^*} =\sum_{p\neq p'} \frac{\omega_{pp'}}{\hbar}(f_{p'}-f_p)|\bra{p'}\hat{R}_x\ket{p}|^2.
\end{equation}
Inserting this into the previous expression, one can show that, in the low $\Gamma$ limit, Eq.~\eqref{eq:Kubo} and Eq.~\eqref{eq:conductivity} are equivalent given a single transition frequency $\omega_{pp'}=\pm\sqrt{m/m^*}\omega_0$. Thus, in the harmonic and weakly-interacting limit, the two analyses converge with $\Gamma=\tau_Q^{-1}$.

A second comparison can be made for arbitrary $\Gamma$ near resonance. When $\omega \approx \sqrt{m/m^*}\omega_0$, we can rewrite Eq.~\eqref{eq:conductivity} as 
\begin{equation} \label{eq:conductivity2} 
\sigma(\omega) = \frac{Ni\omega}{2m^*\sqrt{m/m^*}\omega_0}\frac{1}{\omega - \sqrt{m/m^*}\omega_0 + i\Gamma/2}.
\end{equation} 
This again has a similar form to Eq.~\eqref{eq:Kubo} with a single resonant frequency $\omega_{pp'} = \sqrt{m/m^*}\omega_0$.

\clearpage
\onecolumngrid
\appendixpageoff
\appendixtitleoff
\begin{appendices}
\resumetoc 

\begin{center}
\huge
Supplemental Material
\normalsize
\end{center}

\setcounter{equation}{0}
\setcounter{figure}{0}
\setcounter{table}{0}
\setcounter{page}{1}
\makeatletter
\renewcommand{\theequation}{S\arabic{equation}}
\renewcommand{\thefigure}{S\arabic{figure}}
\renewcommand{\thesection}{S\arabic{section}}


\tableofcontents

\section{2-body T-matrix in a lattice}\label{sec:T2}

We start from the general definition of the $\mathcal{T}$-matrix operator (we take $a_L=1$ in this section): 
$$\hat{\mathcal{T}}=\hat{V}+\hat{V}\hat{G}_0\hat{\mathcal{T}}.$$ 
We take matrix elements between two-body Bloch states $|\bp'_1,\bp'_2\rangle$ and $|\bp_1,\bp_2\rangle$ and insert a closure relation:
$$\langle \bp_1,\bp_2|\hat{\mathcal{T}}|\bp'_1,\bp'_2\rangle=\langle \bp_1,\bp_2|\hat{V}|\bp'_1,\bp'_2\rangle
+\sum_{\bp''_1,\bp''_2}\langle \bp_1,\bp_2|\hat{V}|\bp''_1,\bp''_2\rangle \langle \bp''_1,\bp''_2|\hat{G}_0
|\bp''_1,\bp''_2\rangle\langle \bp''_1,\bp''_2|\hat{\mathcal{T}}|\bp'_1,\bp'_2\rangle
$$\\
For the on site two-body interaction, we have $\langle \bp_1,\bp_2|\hat{V}|\bp'_1,\bp'_2\rangle=U/N_\mathrm{s}\,\tilde{\delta}_{\bp_1+\bp_2,\bp'_1+\bp'_2}$, where $\tilde{\delta}$ is nonzero iff $\bp_1+\bp_2=\bp'_1+\bp'_2+\bG$, where $\bG$ is a vector of the reciprocal lattice. We find
$$
\langle \bp_1,\bp_2|\hat{\mathcal{T}}|\bp'_1,\bp'_2\rangle
=\frac{U}{N_\mathrm{s}}\,
\tilde{\delta}_{\bp_1+\bp_2,\bp'_1+\bp'_2}
+\sum_{\bp''_1,\bp''_2}\frac{U}{N_\mathrm{s}}\,
\tilde{\delta}_{\bp_1+\bp_2,\bp''_1+\bp''_2}
\frac{1}{z-\varepsilon_{\bp''_1}-\varepsilon_{\bp''_2}}
\langle \bp''_1,\bp''_2|\hat{\mathcal{T}}|\bp'_1,\bp'_2\rangle
$$\\
By iteration of this equation, we find that the sum of the momenta of the two particles is conserved modulo a vector of the reciprocal lattice at each order, and we find $
\langle \bp_1,\bp_2|\hat{\mathcal{T}}|\bp'_1,\bp'_2\rangle
=\tilde{\delta}_{\bp_1+\bp_2,\bp'_1+\bp'_2}\mathcal{T}(\bp_1+\bp_2;z)/N_\mathrm{s}$. 
The equation for $\mathcal{T}$ is an algebraic one and we find
\begin{equation}
\mathcal{T}(\bP;z)^{-1}=U^{-1}-\TD(\bP,z)\label{eq:eqT2final},
\quad \mbox{where} \quad 
\TD(\bP,z)=\frac{1}{N_\mathrm{s}}\sum_{\bq_1,\bq_2}\frac{\tilde{\delta}_{\bP,\bq_1+\bq_2}}{z-\varepsilon(\bq_1)-\varepsilon(\bq_2)} =\frac{1}{N_\mathrm{s}}\sum_{\bq}\frac{1}{z-\varepsilon(\bq)-\varepsilon(\bP-\bq)} \,.
\end{equation}
where the second step follows from $\varepsilon(\bP-\bq_1+\bG)=\varepsilon(\bP-\bq_1)$ for any vector $\bG$ of the reciprocal lattice.
To calculate $\TD(\bP,z)$, we make the change of variable $\bq=\bq'+\bP/2$. 
In the $N_\mathrm{s}\to\infty$ limit, $N_\mathrm{s}^{-1} \sum_{\bq'\in 1BZ}$ can be replaced by $\int\! d^3 q'/(2\pi)^3$. 
We use a tight-binding dispersion, such that
\begin{equation} \label{eq:esum}
\varepsilon(\bq)+\varepsilon(\bP - \bq)=-2t\big[\sum_{\alpha}\cos(q'_{\alpha}+P_{\alpha}/2)+\cos(P_{\alpha}/2-q'_{\alpha})\big]=-4t \sum_{\alpha}\cos(P_{\alpha}/2)\cos(q'_{\alpha})\,,\end{equation} 
where $\alpha=\{x,y,z\}$ is the index of directions.
So far we have 
\begin{equation} \label{eq:Pi_int_final}
\TD(\bP,z)=\int\! \frac{d^3q}{(2\pi)^3} \, 
\frac{1}{z+4t\sum_{\alpha}\cos(P_{\alpha}/2)\cos(q_{\alpha})}
\end{equation}
This expression can be written using 
the identity $1/Z=-i\int_0^{+\infty}\! du \, e^{i Z u}$ for $\Im Z>0$. 
For $Z=z-\varepsilon(\bq)-\varepsilon(\bP - \bq)$ written as Eq.~\eqref{eq:esum}, this gives
$$
\frac{1}{z-\varepsilon(\bq)-\varepsilon(\bP - \bq)}=-i\int_0^{+\infty} \! du \, e^{i z u} \prod_{\alpha}
\exp[4it\,\cos(P_{\alpha}/2)\cos(q'_{\alpha})u]\, .
$$
In the calculation of 
$\TD(\bP,z)$, the integrals on the $q'_{\alpha}$'s factorize. We use the identity
$$
J_0(x)=\int_{-\pi}^{\pi}\! \frac{dq}{2\pi} \, e^{i\,x\,\cos(q)}
.$$
In this way, we find
\begin{eqnarray}
\TD(\bP,z)&=&-i\int_0^{+\infty}\! du \, e^{i z u}\prod_{\alpha}J_0[4t\,\cos(P_{\alpha}/2) u]
\end{eqnarray}
In practice we perform this integral numerically, which converges absolutely in our 3D problem. Using Eq.~\eqref{eq:eqT2final}, we obtain the $\mathcal{T}$-matrix.

\section{Further information on the dissipation model} 

\subsection{Calculation of moments\label{app:momentscal}}

Here we derive Eqs.~(2) and (3) of the main text and then their different terms, called moments, using the ansatz from Eq.~(5), which leads to Eqs.~(6) and (7). By ``moments" we mean ensemble-average of single-particle observables,   
$\langle \mathcal{O} \rangle=\int \mathcal{O} f(\bp,\br,\ttime)/N_{\sigma}$, where $\int$ denotes $\int\! d^3r\!\int \! d^3p/(2\pi\hbar)^3$, the momentum integral is restricted to the first Brillouin zone, and $N_{\sigma}$ is the number of particles per spin.

To get Eq.~(2), we start with the time derivative of position 
$d\langle \br\rangle /d\ttime=\int\! \br\, \partial_\ttime f(\bp,\br,\ttime)/N_{\sigma}$. 
Consider the three contributions to $\partial_\ttime f$ in Eq.~(1). 
The first term (proportional to $\bv_{\bp}$) gives, after integration by parts, the expectation value $\langle \bv\rangle$. The second term (proportional to $\bm F$) cancels out after integration by parts with respect to momentum. 
The contribution of the collision integral  $I_{\rm coll}$ also vanishes since collisions are local. We thus find the expected result, 
$d \langle \br\rangle/d\ttime=\langle\bv\rangle$. 
In the same manner, we can find Eq.~(3), for the time derivative of the velocity. Integration on the position of the first term gives zero. The second term, for the $\alpha$  component of $\bv$, after integration by parts on momentum, yields 
$\langle F_{\beta}m^{-1}_{\alpha\beta}\rangle$,
where we have introduced the inverse effective-mass matrix 
$m^{-1}_{\alpha\beta}=\partial v_{\bp,\alpha}/\partial p_{\beta}=\partial^2 \varepsilon_{\bp}/\partial p_{\beta}\partial p_{\alpha}$. The last term, involving the collision integral cannot be simplified in general.

We now compute the different terms entering Eqs.~(2) and (3).

In the calculation of $\langle \br\rangle$, we make the changes of variables $\bp'=\bp-\bq(\ttime)$ and $\br'=\br-\bR(\ttime)$ and we find
\begin{equation} \langle \br\rangle=\bR(\ttime) \end{equation} 
This is valid for any $\bR(\ttime)$.
Taking the total time derivative of $\langle \br\rangle$ and using Eq.~(1), we obtain for vanishing $\bq(\ttime)$ and $\bR(\ttime)$: 
$$\langle v_{\alpha}\rangle=N_{\sigma}^{-1} \int\! \Big(-\frac{\partial f^0}{\partial E}\Big) v_{ p,\alpha}
\big(\bv\cdot\bq(\ttime)+m\,\omega_0^2\,\br\cdot\bR(\ttime)\big) \, .$$
The second term gives zero after integration on $p_{\alpha}$, since it is an odd function of $p_{\alpha}$.
The first term gives
\begin{equation} \langle v_{\alpha}\rangle=A_{\alpha,\beta}q_{\beta}(\ttime) \end{equation}
where $A_{\alpha,\beta}=N_{\sigma}^{-1} \int\! (-\partial f^0/\partial E)v_{\bp,\alpha}v_{\bp,\beta}$. An integration by parts with respect to $p_{\alpha}$ enables to find that $A$ is simply related to the effective mass
\begin{equation} \label{eq:A}
A_{\alpha,\beta}= N_{\sigma}^{-1} \int\!  \frac{\partial^2\,\varepsilon_{\bp}}{\partial p_{\alpha}\partial p_{\beta}}f^0(\bp,\br)=\langle  m^{-1}_{\alpha\beta}\rangle_{\mathrm{eq}} \end{equation}
where the $\langle \cdot \rangle_\mathrm{eq}$ average is performed with $f^0$.
For a cubic lattice, $A_{\alpha,\beta}=A\,\delta_{\alpha,\beta}$.

The collision integral from the Boltzmann equation is 
\begin{equation} \label{eq:Icoll} I_{\rm coll}[f](\bp_1,\br,\ttime)=-\int\!\frac{d^3p_2 \, d^3p_3}{(2\pi\hbar)^6}\,\Gamma_{12,34} \\ \big[f_1\,f_2(1-f_3)\,(1-f_4)-f_3\,f_4\,(1-f_1)\,(1-f_2)\big] \end{equation}
where $f_i \equiv f(\bp_i,\br,\ttime)$. $\bp_4$ is determined by momentum conservation as $\bp_4=\bp_1+\bp_2-\bp_3+\bG$, where $\bG$ is a vector of the reciprocal lattice such that all momenta stay in the first Brillouin zone. 
The scattering rate at which $\{\bp_1,\bp_2\}\to\{\bp_3,\bp_4\}$ is given by the generalized Fermi golden rule
\begin{eqnarray}
\Gamma_{12,34}&=&\frac{2\pi}{\hbar}\delta(E_{12}-E_{34})|\langle \bp_3\,\bp_4 |\hat{\mathcal{T}}(E_{12})|\bp_1\,\bp_2\rangle |^2,
\end{eqnarray}
where $E_{ij}=\varepsilon_{\bp_i}+\varepsilon_{\bp_j}$ is the energy of two particles of quasimomenta $\bp_i$ and $\bp_j$.

To evaluate the collision integral, we define the new function $\varphi$ such that
\begin{equation} \label{eq:fdeviation} 
f(\bp,\br,\ttime)=f^0(\bp,\br)
-\frac{\partial f^0}{\partial E}\varphi(\bp,\br,\ttime)+\cdots \end{equation}
In the ansatz we have 
\begin{equation}  \varphi(\bp,\br,\ttime)=\bv\cdot\bq(\ttime)+m\,\omega_0^2\,\br\cdot\bR(\ttime).\label{eq:Ansatzphi} \end{equation}
Following a standard procedure, we linearize the Boltzmann equation. Using that the Fermi-Dirac distribution is the equilibrium ($I_\mathrm{coll}[f^0]=0$), we substitute Eq.~(5) in the collision integral to get, at first order,
\begin{equation} \label{eq:Icollkinetic} 
I_\mathrm{coll}[f](\bp,\br,\ttime)=
-\frac{1}{T}\int\,\Gamma_{12,34}
f_1^0\,f_2^0(1-f_3^0)(1-f_4^0)\big(
\varphi_1+\varphi_2-\varphi_3-\varphi_4
\big)
+\cdots \end{equation}
In the integral $N_{\sigma}^{-1} \int\! \bv_{\bp}I_\mathrm{coll}[f]$, that is local in space, the second term of Eq.~\eqref{eq:Ansatzphi}, which is independent of $\bp$, does not contribute since the integral on ${\bf p}$ vanishes by parity. We write the scalar product $\bv\cdot\bq(\ttime)=v_{p,\alpha}\,q_{\alpha}(\ttime)$ and we find 
\begin{equation}
N_{\sigma}^{-1} \int\! v_{\bp,\alpha}I_\mathrm{coll}[f]=-B_{\alpha,\beta}q_{\beta},
\end{equation}
where $B_{\alpha,\beta}$ is given by Eq.~(6) of the main text. 
For a cubic lattice, $B_{\alpha,\beta}=B\,\delta_{\alpha,\beta}$.

As the total force is the sum of the trap confinement and the driving force, we have $\langle F_{\beta}m^{-1}_{\alpha\beta}\rangle=-m\,\omega_0^2\,\langle\,r_{\beta} m^{-1}_{\alpha\beta}\rangle+F_0\,\cos(\omega \ttime)\langle m^{-1}_{\alpha\beta}\rangle\delta_{\beta,x}$. At first order, the second term (for a cubic lattice and a separable potential) is $F_0\,\cos(\omega \ttime)\langle 1/m^*(p_x)\rangle_{\mathrm{eq}} \delta_{\beta,x}$. 
The first term gives zero at lowest order, after integration on $r_{\alpha}$. The first order contribution is given by
\begin{equation}
N_{\sigma}^{-1} \int \Big(-\frac{\partial f^0}{\partial E} \Big) \big(\bv\cdot\bq(\ttime)+m\,\omega_0^2\,\br\cdot\bR(\ttime)\big)\,r_{\beta}
 m^{-1}_{\alpha\beta}\nonumber
\end{equation}
The integration on $r_{\beta}$ makes the first term vanish, since the integrand is an odd function of $r_{\beta}$. We rewrite the second term of the integral as $-m\,\omega_0^2\,\langle\,r_{\beta}  m^{-1}_{\alpha\beta}\rangle\equiv-C_{\alpha,\beta}\,R_{x\beta}$
where
\begin{equation}
C_{\alpha,\beta}=\frac{(m\omega_0^2)^2 }{N_{\sigma}} \int \Big(-\frac{\partial f^0}{\partial E} \Big)r_{\beta}r_{\gamma}
 m^{-1}_{\alpha\,\gamma} \,.
\end{equation}
We integrate by parts with respect to the coordinate $x$, keeping all other variables fixed.
In the integration by parts, we take $U'= (-\frac{\partial f^0}{\partial E})\,m\,\omega_0^2\,x=(-\frac{\partial f^0}{\partial E})\,\partial_x E=-\partial_x f^0$ and 
$V=x$: $\int_{-\infty}^{+\infty} (-\frac{\partial f^0}{\partial E})\,m\,\omega_0^2\,x^2\,dx=[-f^0\,x]_{-\infty}^{+\infty}+\int_{-\infty}^{+\infty} f^0\,dx=\int_{-\infty}^{+\infty}\,f^0\,dx$.
As a consequence, we find at lowest order
\begin{equation} -m\,\omega_0^2\,\langle\,r_{\beta}  m^{-1}_{\alpha\beta}\rangle
 =-m\,\omega_0^2\langle m^{-1}_{\alpha\beta}\rangle_{\mathrm{eq}}\,R_{\beta} 
 =-\frac{m}{m^*}\,\omega_0^2\,R_{\alpha}
\end{equation}
where we have used that 
$\langle m^{-1}_{\alpha\beta}\rangle_{\mathrm{eq}}=\delta_{\alpha\beta}/m^*$ for a cubic lattice.

\subsection{Relation of collision rate to current dissipation rate}

In the case where the perturbation is purely a displacement in quasi-momentum, relaxation of the distribution involves only those events that change the net current. One can show this as follows. The collisional relaxation rate of the distribution function is given by Eq.~\eqref{eq:Icollkinetic}. 
For a perturbation of the form \eqref{eq:fdeviation}, one can write
$\varphi_1+\varphi_2-\varphi_3-\varphi_4$ as $-\Delta p_\beta \Delta J_\beta(12;34)$, 
where $\Delta J_\beta(12;34) = v_{ p_1,\beta}+v_{p_2,\beta}-v_{p_3,\beta}-v_{p_4,\beta}$. 
Thus $\partial f_1 / \partial \ttime$ involves a sum over collisional events, each weighted by $\Delta J_\beta(12;34)$. Those scattering events which do not change the current have $\Delta J_\beta(12;34)=0$ and do not contribute to the relaxation rate. An example of such events are those in which all momenta remain in the quadratic part of the dispersion relation. 

\subsection{Density dependence of dissipation rate \label{sm:density}}

\begin{figure}[tb!]
\includegraphics[width=0.5 \columnwidth]{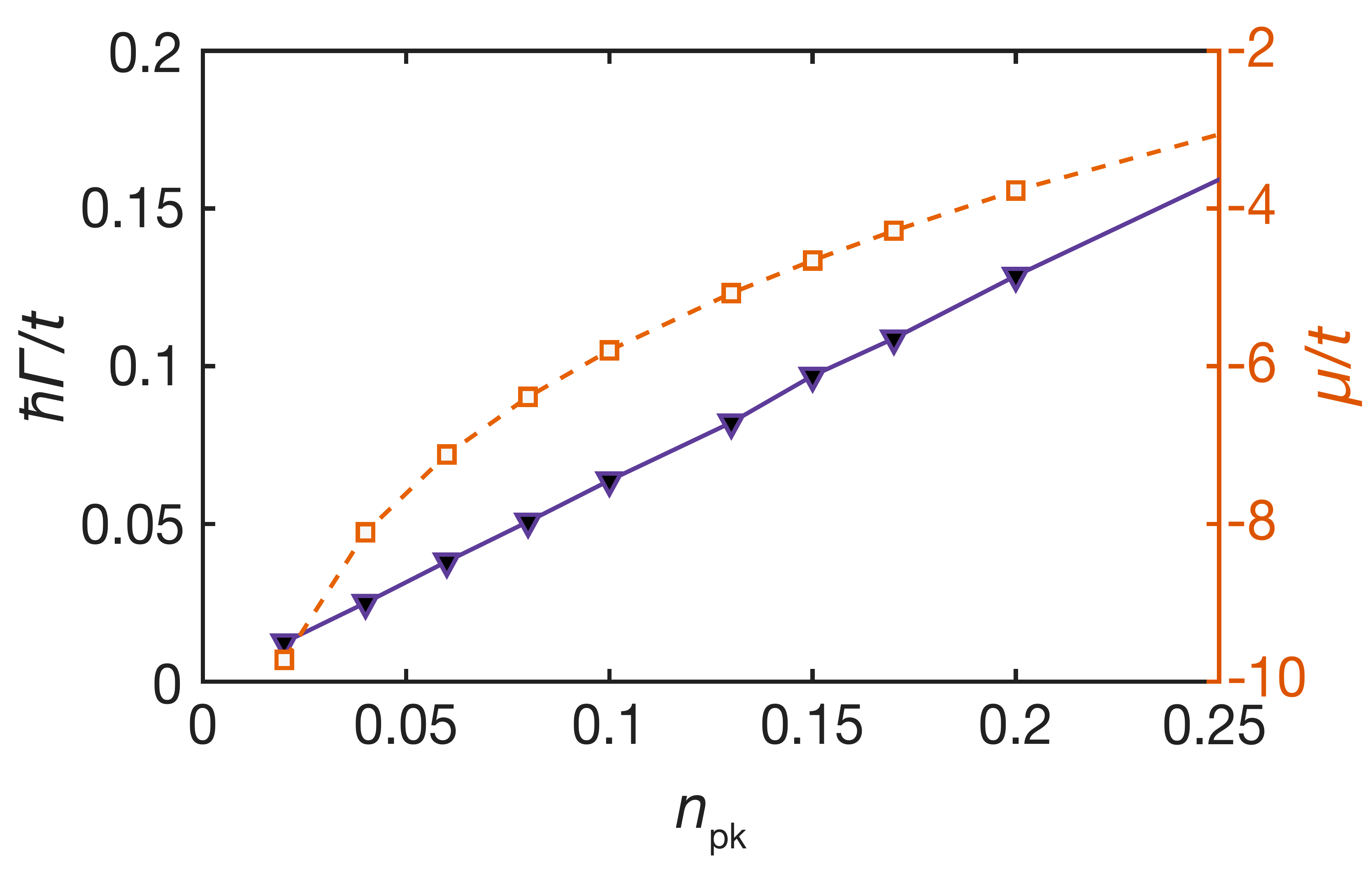}
\centering
\caption{\label{fig:density} {\bf Density dependence of $\Gamma$.} The current dissipation rate (purple triangles) and chemical potential (orange squares) is calculated for $U/t = 5.75$ and $T/t=2.19$ versus the peak filling.}
\end{figure}

Figure \ref{fig:density} shows the  scattering rate $\Gamma$ calculated by the dissipation model across a range in peak filling, for typical experimental parameters, $U/t = 5.75$ and $T/t = 2.19$. The dissipation rate linearly increases with density at low fillings. Indeed at these fillings and temperature the Fermi-Dirac distribution is approximately equal to the Maxwell-Boltzmann one, so is proportional to the fugacity $e^{\beta \mu}$, and the filling $n \simeq \int d^3\bm{k} \exp \beta (\mu-\varepsilon_{\bm{k}})/(2\pi)^3 \simeq e^{\beta \mu} I_0(\beta t)$ where $I_0$ denotes the Bessel integral of the first kind, is also proportional to the fugacity. With a Maxwell-Boltzmann distribution, $B$ of Eq.~(6) is then proportional to $n^2$ and $m^*$ to $1/n$, so the scattering rate $\Gamma=Bm^*$ to $n$. We notice that the curve of Fig.~\ref{fig:density} is less linear at high fillings, where the Maxwell-Boltzmann approximation breaks down.

\subsection{Second- and third-order tunneling}

In this supplement we want to quantify the effect of higher order hopping on our analytical model.
The trapping potential due to the lasers is $V_\mathrm{opt}({\bf r})=V_0\sum_{\alpha}\sin(k_Lx_{\alpha})^2=V_0/2\sum_{\alpha}(1-\cos(2k_Lx_{\alpha})\big)$. As it is separable, we write the corresponding Schr\"{o}dinger equation only in the $x$ direction:
\begin{equation}
-\frac{\hbar^2}{2m}\frac{d^2\Psi}{dx^2}-\frac{V_0}{2}\cos(2k_Lx)\Psi=E\Psi
\end{equation}
With the substitutions $\tau=k_Lx$ and $q=-mV_0/(2\hbar^2k_L^2)=-V_0/(4E_R)$, one
recognizes the Mathieu equation,
\begin{equation}
f''+[a-2q\cos{(2\tau)}]f=0 \, ,
\end{equation}
whose solutions have been studied extensively.

The formal solution can be Fourier developed to next-next-nearest (3NN) neighbor:
\begin{equation}
E(k)=E_0-2t\cos(ka_L)-2t'\cos(2ka_L)-2t''\cos(3ka_L)+\cdots
\end{equation}
with the numerical values $t'/t = -0.12011$, $t''/t = 0.02378$ for the 2.5\,$E_R$ lattice depth used in the experiment.

This dispersion relation including up to the $t'$ term can be implemented in our calculation of $\Gamma$ through the $\TD$ term of $\mathcal{T}$-matrix. For this we come back to Eq.~\eqref{eq:eqT2final}, but now cannot reduce it to an integral of Bessel functions. Instead, one can use the identity
\begin{equation}
\int \frac{1}{g(x)+i0^+} = \mathcal{P}\mathcal{V} \int \frac{1}{g(x)} - i \pi \int \delta(g(x)) \, ,
\label{eq:Cauchy}
\end{equation}
where $\mathcal{P}\mathcal{V}$ denotes the Cauchy's principal value of an integral. Then, numerically, we replace the Dirac delta of the imaginary part by a `door' function, and the principal value by an `anti-door' function, being zero if $|g(x)|<\sigma_0$ and $1/g(x)$ else.

\begin{figure}[tb!]
\includegraphics[width=0.8 \columnwidth]{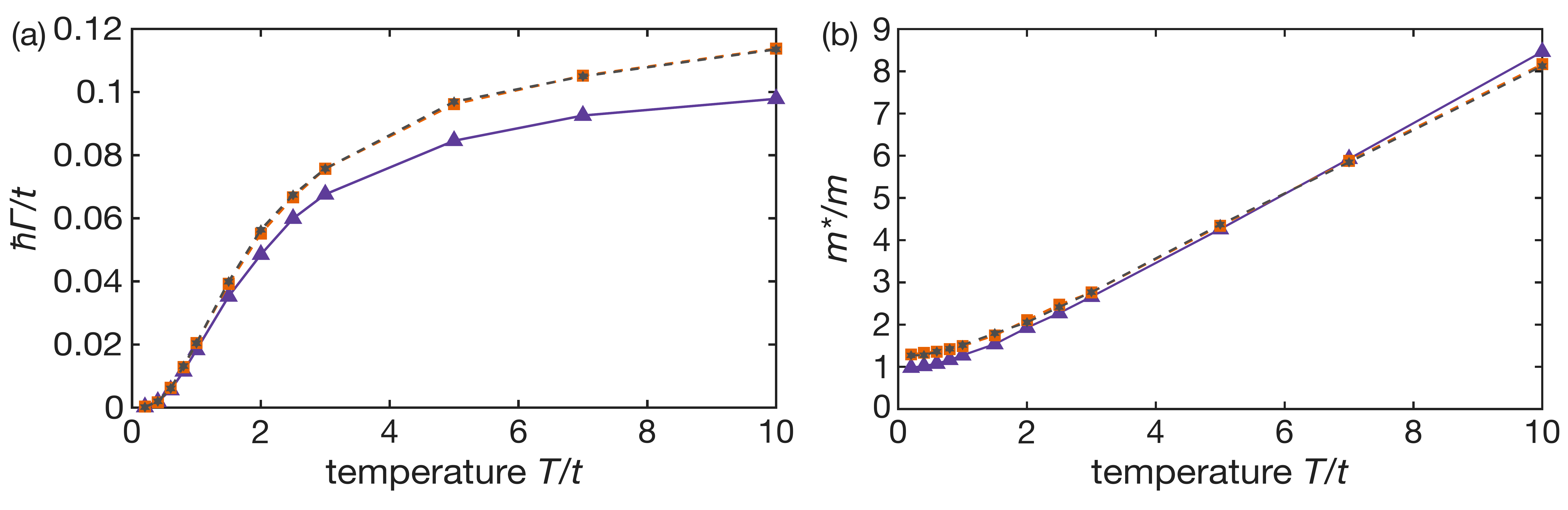}
\centering
\caption{\label{fig:FigS2} {\bf Effect of next-nearest neighbor tunneling. (a)} The current dissipation rate, $\hbar \Gamma/t$, is calculated for a range of temperatures for the same conditions as Fig. 4 using a tight-binding approximation (purple triangles), including next-nearest neighbor tunneling (orange squares), and next-next-nearest neighbor tunneling (gray hexagrams). The difference between the tight-binding and next-nearest for all $T < 10t$ is around $12\%$. Third-order tunneling is within $2\%$ of the next-nearest neighbor calculation. {\bf{(b)}} The effective mass is calculated using Eq. \ref{eq:A} for a tight-binding approximation (purple triangles), including next-nearest neighbor tunneling (orange squares), and next-next-nearest neighbor tunneling (gray hexagrams). We see that corrections beyond next-nearest neighbor are negligible.} 
\end{figure}

In this way, one can implement the new dispersion relation in the expression for the current dissipation $\Gamma$, and compare to the next-nearest neighbor approximation. The result is shown in Fig.~\ref{fig:FigS2}(a) for the parameters of Fig.~4: $U/t = 5.75$ and $n_{\mathrm{pk}} = 0.095$. For $\Gamma$, we find that the tight-binding and next-nearest neighbor approximations differ by over 10\%. The next-next-nearest neighbor effect on $\Gamma$ is within $2\%$ across the range. Since higher-order tunneling is significant, we choose to use up to the $t''$ term in the calculations of $\Gamma$ in the main text. 

Figure~\ref{fig:FigS2}(b) shows $m^*/m$ versus $T$ for the same parameters. We see that the tight-binding and next-nearest neighbor calculation differs by as much as 20\% at some temperatures. The third order term involving $t''$ agrees within 3\% across the range of $0<T/t<10$. For this reason we calculate $m^*$ using up to the $t''$ term in the main text. 

\section{Further information on experimental methods } 

\subsection{In situ imaging protocol \label{sm:microscopy2}} 

The in situ charge distribution is measured via quantum gas microscopy, and in this section we discuss notable upgrades to the imaging and measurement protocol as described in our prior work. After applying the time-varying force, a 70\,$E_R$ pinning lattice is turned on in \qty{200}{\us} to freeze the atomic motion. Then the magnetic field is ramped to 4\,G, and a near resonant laser pulse optically pumps all atoms into the $a$ state. All $ab$ doublons are ejected from the trap via photoassociation and subsequently imaged as holes in the fluorescence distribution. 

The central $xy$ plane of the three-dimensional lattice is selected for imaging via microwave spectroscopy. A 650\,G/cm field gradient and a \qty{124}{\gauss} bias field are applied to magnetically separate each plane by \qty{34}{\milli\gauss}. The desired plane to image is shelved in the $\ket{F,m_F} = \ket{7/2,-7/2}$ state via an HS1 microwave sweep, and the undesired planes are ejected from the trap via a $\sim\qty{1}{\ms}$ resonant laser pulse. A large bias coil in conjunction with a FL1-100 Fluxgate magnetometer is used to shim out slow drifts in the ambient laboratory field at the $<\qty{30}{\milli\gauss}$ level. Slow mechanical drifts of the position of the electromagnets relative to the center plane induce field drifts of $\sim 100$\,mG on the multi-hour time scale, and thus the selected plane is further stabilized by applying a tilted bias field during microwave spectroscopy to image multiple planes as ``stripes". The position of the stripes is proportional to the magnetic field, allowing us to feedback back onto the selection frequency between measurements of $\sigma(\omega)$. After a three-hour warm-up period, the same vertical plane is selected for imaging for $>120$ hours during continuous experiment operation. 

\begin{figure}[tb]
    \centering    \includegraphics[width=0.7\linewidth]{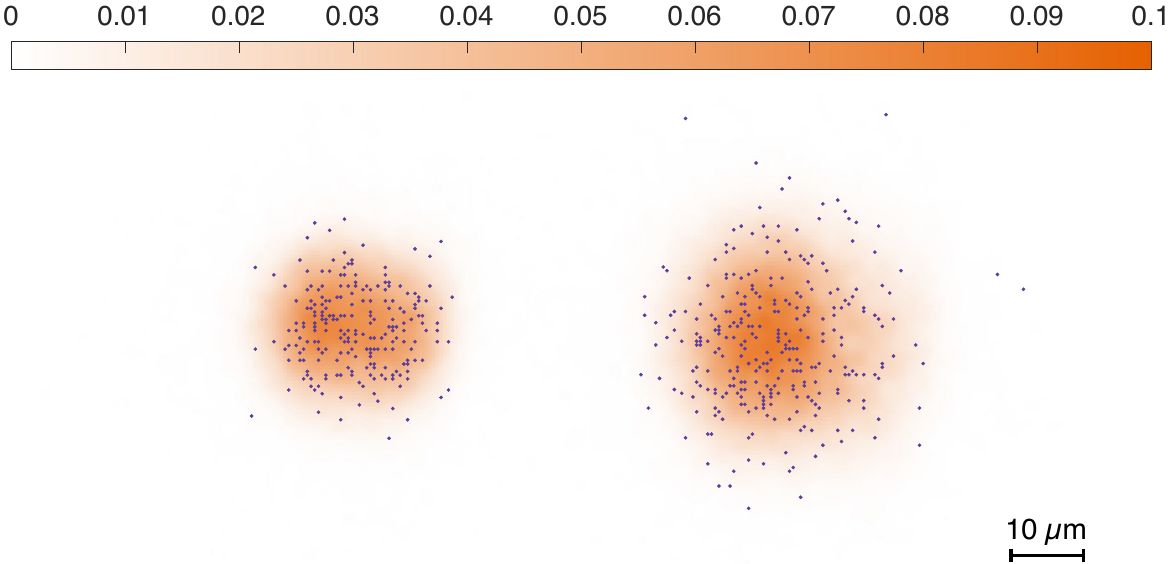}
\caption{{\bf In situ atomic distribution.} Example in situ digitized images at $T/t = 1.2(2)$ (left) and $T/t = 2.6(4)$ (right). The orange atomic clouds represent the average filling $n(\bf{r})$ of $\sim$ 300 digitized images, smoothed by a Gaussian filter with a binning size of 1.5 lattice sites. The purple atoms show a single digitized image randomly selected from the set of images used in the averaged distributions. } 
    \label{fig:qgm}
\end{figure}

After selecting the desired plane to image, the atomic spatial distribution is measured by capturing the fluorescence from combined electromagnetically induced transparency (EIT) cooling and Raman sideband cooling (RSC) over \qty{4}{\s}. The Richardson-Lucy algorithm is utilized to sharpen the image with the point spread function of the microscope objective, and the underlying lattice structure and phase is reconstructed by analyzing the sharpened image's Fourier distribution. A spatially dependent threshold is applied to identify occupied lattice sites. Example digitized images are shown in Fig. \ref{fig:qgm}. Between measurements of $\sigma(\omega)$, the imaging fidelity is measured by comparing the fluorescence of successive \qty{2}{\s} exposures. The typical extracted hopping (loss) rate is estimated to be 7\% (12\%). The location of the microscope objective focus is also optimized between measurements of $\sigma(\omega)$.

In the low density limit, and with a balanced mixture, the directly observed charge filling is twice the filling $n(\bf{r})$ per spin state. To account for loss during in situ imaging we divide the observed filling by the measured fidelity.
The peak filling $n_{\mathrm{pk}}$ is calculated for each image using the number of atoms and bootstrapped second moment, assuming a gaussian distribution. 
For the data shown in Figs.~3 and 4, the distribution is well-approximated by the Boltzmann high-temperature limit, where the volume-averaged filling, $n$, is $\sqrt{8}$ smaller than $n_{\mathrm{pk}}$. For each spectrum, $n_{\mathrm{pk}}$ is averaged over $\sim300$ images.

\subsection{Thermometry \label{sm:microscopy}}

Temperature, $T$, is measured by comparing in-situ filling to a thermal model. A typical sequence includes a 100\,ms ramp+hold pre-thermalization drive described in the main text, followed by imaging (see \S\ref{sm:microscopy2}). Since our experiments are conducted in a high-temperature, low filling regime, one expects that a Maxwell-Boltzmann (MB) distribution would give a valid interpretation of the in situ distribution. In this section, we describe how $T$ is determined and estimate corrections to MB thermometry due to FD statistics and finite interaction strength. 

In the $t \to 0$ limit, the Hubbard Hamiltonian is the sum of purely on-site Hamiltonians: 
\begin{equation} \label{eq:FHMtzero}
\hat H_U -\mu \hat N = \sum_j \hat h_{j}, 
\qquad \mbox{with} \quad 
\hat h_{j} = U \hat{n}_{j,\uparrow} \hat{n}_{j,\downarrow} - \mu (\hat{n}_{j,\uparrow} +\hat{n}_{j,\downarrow}) \end{equation} 
Each site has four possible configurations: no atoms, singlon $\upa$, singlon $\dna$, and doublon occupation. 
If we assume that all sites have a common $\beta$ and $\mu$, each of these has a probability proportional to the Gibbs factor $\exp(-\beta E + \beta \mu N)$. Setting the lowest band energy to zero, the Gibbs factors for the four possible configurations are now $z^2 \exp(-\beta U)$ for a doublon, $z$ for either singlon, and $1$ for empty occupancy, where $z = \exp(\beta \mu)$. Normalizing by the partition function $\mathcal{Z}$:
\begin{equation} \label{eq:Gibbs}
P_\mathrm{zero} =  \mathcal{Z}^{-1} 
 \quad \mbox{,~} \quad
P_\mathrm{single} =  2 z \mathcal{Z}^{-1}
 \quad \mbox{,~} \quad
 P_\mathrm{doublon} = z^2 \mathcal{Z}^{-1} e^{-\beta U}
 \quad \mbox{with} \quad
  \mathcal{Z} = z^2 e^{-\beta U} + 2 z + 1
\end{equation}

The MB limit can be found by expanding these probabilities in the $z \ll 1$ limit. At lowest order, $P_\mathrm{single} \approx 2 z$. Using the local-$\mu$ approximation, that $\mu(r) = \mu_0 - V(r)$, we then have 
\begin{equation} \label{eq:Boltzmann}
P_\mathrm{single} \approx 2 z_0 \exp[-\beta V(r)] 
\quad \mbox{with} \quad \
z_0 = \exp[-\beta \mu_0] 
\qquad \mbox{(MB limit)}
\end{equation}
We calculate the average second moment of the imaged filling distribution for each spectrum ($\sim 300$ images) and use Eq.~\eqref{eq:Boltzmann} to calculate $\beta$. $T$ is taken from the $y$-lattice direction, perpendicular to the drive direction, as Joule heating and localization can inflate the calculated equilibrium temperature. We independently calibrate the trap frequency in $V(y)$ by displacing a non-interacting ensemble and observing the oscillation frequency. 

For a fugacity of $z=0.2$, the complete Gibbs estimate gives a parity-projected filling of $n_\mathrm{pk} \approx 0.3$, whereas the Boltzmann estimate gives
a parity-projected $n_\mathrm{pk} \approx 0.4$. We also note that the effect of interactions is small (of order $\leq 0.01$ correction for $z \leq 0.4$), because the probability of doublons is low. In a power-law expansion in $z$, $P_\mathrm{single} = \sum_{i=1} c_i z^i$, interactions appear at third order, $c_3 = 8 - 2 e^{-\beta U}$, and the leading order correction to the observed filling is instead due to Pauli blocking: $c_2 = -4$. 

\subsection{Uncertainty estimation for temperature \label{sm:tempuncertainty}}

The uncertainties in the reported $T$ are estimated by combining the uncertainty in the trap frequency and the statistical uncertainty in the average second moment of the atomic distribution for each spectrum. 
The uncertainty in the trap frequency is determined by comparing the best-fit trap frequency from two independent measurements: (A) observation of the oscillation frequency after an initial displacement of a non-interacting ensemble and (B) a Kubo-model fit of the conductivity spectra (see End Matter), in which $\omega_0$ is a free parameter. We take the standard deviation of the difference between the methods, here \qty{2.6}{\Hz}, as the uncertainty in trap frequency. Since these measurements were made on separate days, this also captures the typical effect of alignment drift. 
The statistical uncertainty in the average second moment is obtained in a fairly standard manner: standard error of the mean second moment, each of which is obtained via bootstrap analysis of $\sim300$ in-situ images for a given spectrum. 
All reported temperature uncertainties are given by twice this calculated standard error (95\% confidence). 

\section{Further information on resistivity analyses}

\begin{figure}[tb!]
\includegraphics[width=0.8 \columnwidth]{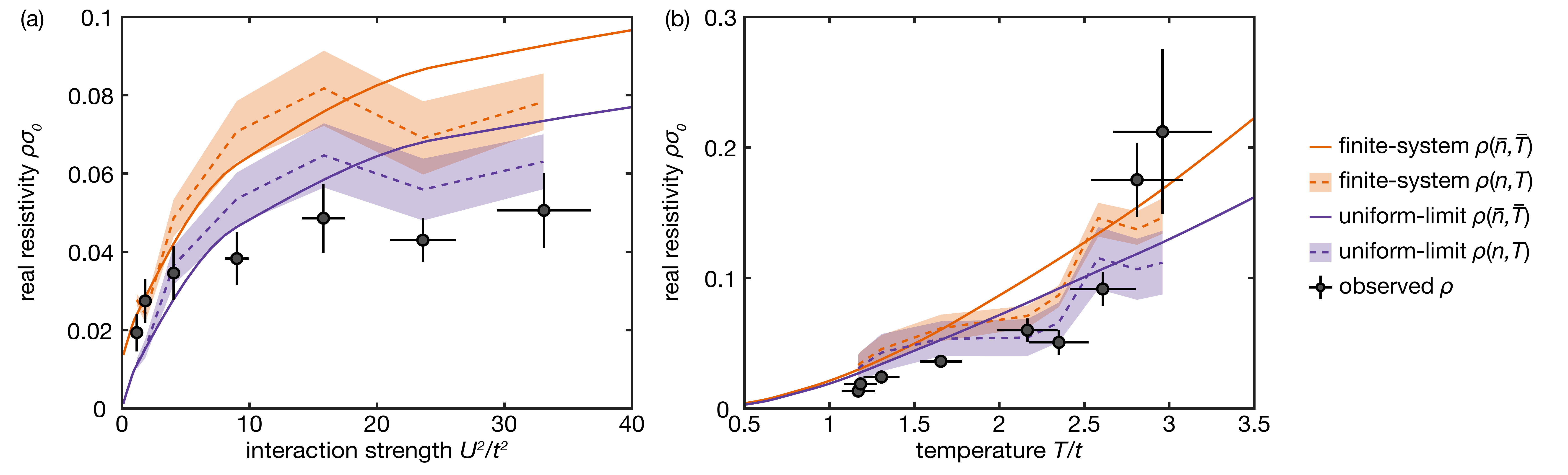}
\centering
\caption{\label{fig:modelComp} {\bf Uniform-limit versus finite-system resistivity.} 
{\bf (a)} For each $U$, the calculated $\Gamma$ is converted to $\Re \rho$ using two alternatives: 
the uniform-limit choice (shown in purple) uses the kinetic model, Eq.~(8) from the main text; and the finite-system choice uses the Kubo fit model (shown in orange), Eq.~(9), with $\tau_Q^{-1} = \Gamma$, at the frequency $\omega$ where $\Im \rho = 0$. As discussed in the text, these differ due to the anharmonicity of a finite system. For each alternative, a trend line calculated for the average $n$ and $T$ across all runs (solid line), the resistivity calculated for the experimental conditions ($n$ and $T$) of each data point individually (dashed line), and the effect of scatter in $n$ and $T$ (uncertainty band) are shown. The shaded uncertainty bands correspond to twice the propagated standard error in $T$ and $n$ for data sets at each $U$. 
Data points are as in Fig.~3 of the main text. {\bf (b)} The same comparison is shown for the temperature dependence, with the solid line calculated using the averaged $n$ across all data sets, the dashed line calculated using the average $n$ for each data point individually, and the uncertainty bands representing twice the propagated standard error of $n$ and $T$ at each $T$. Data points are as in Fig.~4 of the main text.}
\end{figure}

\subsection{Determination of real resistivity}

The observed $\Re \rho$ shown in Fig.~3 and Fig.~4 of the main text are a weighted average of measurements within 60\% of the frequency at which peak conductivity response is observed; weights are the inverse of the statistical variance at each drive frequency. This frequency range was chosen to exclude points where the measured resistivity is almost purely reactive (imaginary), yielding an unreliable measure of $\Re \rho$. An alternate post-selection protocol -- including only points with a conductivity amplitude above 5\% of the peak value -- yielded similar weighted-average values. 

\subsection{Effect of eigenstate anharmonicity} 
\label{sm:Resistivity}

The single-particle eigenspectrum of lattice-plus-parabola potential is harmonic at low energy, with $\omega_0$ replaced by $(m/m^*_0)^{1/2} \omega_0$, where $m^*_0$ is the $q=0$ effective mass. However, it becomes increasingly anharmonic as the energy becomes comparable to the bandwidth. As a result, a finite-$T$ conductivity spectrum has a finite width even as $\tau_Q^{-1} \to 0$. Similarly, the peak resistivity, where $\Im \rho = 0$, remains finite in the weak-scattering limit. 

This leaves an ambiguity in how resistivity is reported given a calculated $\Gamma$ from the dissipation model. The {\em uniform-limit} choice uses Eq.~8, and reports $\Re \rho = (m^*/N) \Gamma$; the {\em finite-system} choice instead calculates the purely real $\rho$ ($\Im \rho = 0$) using the Kubo fit model, $\tau_Q^{-1} = \Gamma$, the measured temperature, and the known trap frequency. These choices are compared in Fig.~\ref{fig:modelComp}. As in the main text, the solid line is for a calculation of $\rho$ that uses a fixed value of $\bar{T}$ and $\bar{n}_{\mathrm{pk}}$. In Fig.~\ref{fig:modelComp}(a), we see that in the $U \to 0$ limit, the uniform-limit resistivity goes to zero, whereas the finite-system resistivity remains finite, due to the anharmonicity effect. For the conditions probed  in our measurements, these choices differ by an offset of $\sim 0.2 \sigma_0^{-1}$.  
The main figures of the manuscript show uniform-limit analyses, since it does not depend on the choice of trapping potential.

Figure~\ref{fig:modelComp} also quantifies the agreement between observed and calculated $\rho$, accounting for variations in temperature and filling. Although we attempted to maintain a constant filling/temperature (when applicable) between datasets, these parameters had some scatter. The dashed lines in Fig.~\ref{fig:modelComp} show the Re $\rho$ calculated for the conditions of each dataset, using each model. The difference between solid and dashed indicates deviations in temperature and filling from the mean. The difference between the data points and the dashed lines indicates agreement with the model. Error bands are calculated in the same manner as the dashed lines, and represent twice the propagated standard error of $n$ and $T$ (95\% confidence).

\end{appendices}
\end{document}